\documentstyle[aps,floats,epsfig]{revtex}

\tighten

\begin{document}

\title{Recent Results from the RICE Experiment at the South Pole
}

\author{I. Kravchenko,$^1$ D. Seckel,$^2$
J. Adams,$^3$ A. Baird,$^3$ S. Churchwell,$^3$ P. Harris,$^3$ S. Seunarine,$^3$
A. Bean,$^4$ D. Besson,$^4$ K. Byleen-Higley,$^4$ S. Chambers,$^4$
J. Drees,$^4$ S. Graham,$^4$ D. McKay,$^4$ J. Meyers,$^4$
L. Perry,$^4$ J. Ralston,$^4$ J. Snow,$^4$ S. Razzaque$^5$ \\
{\it (1) M.I.T. Lab. for
Nuclear Science, Cambridge, MA  02139\\
(2) Bartol Research Institute, U. of Delaware, Newark DE 19716\\
(3) Department of Physics and Astronomy,
Private Bag 4800, U. of Canterbury, Christchurch, New Zealand \\
(4) KU Dept. of Physics and Astronomy, Lawrence KS
66045 \\
(5) Pennsylvania State University, University
Park, PA 16802} 
}

\maketitle

\begin{abstract}
We present a compilation of recent results, submitted to the
2003 International Cosmic Ray Conference (Tsukuba, Japan). These include:
a) Revised Monte Carlo estimates of the
radiofrequency signals produced by electromagnetic showers in ice,
b) an updated
search for ultra-high energy (UHE) neutrinos based on detection 
of radio-wavelength Cherenkov radiation;
such radiation results from neutrino-induced
electromagnetic showers in cold Polar ice, and
c) An {\it in situ} measurement of the index of refraction through
the South Polar firn. 
\end{abstract}

\section{Introduction and Methods}
The Antarctic icecap has provided an extraordinary 
laboratory for a variety of scientific purposes.
The AMANDA, IceCube\cite{IceCube03}, ANITA\cite{ANITA03}, and
RICE
collaborations seek to use the
dense, solid, large-volume, and extraordinarily
transparent\cite{Evans95} (for $\lambda>100~nm$) 
icecap as a neutrino target; the pioneering
AMANDA experiment has 
sucessfully demonstrated the viability of in-ice optical
detection of atmospheric neutrinos, through the reconstruction of
hundreds of muon neutrinos. 
All these experiments seek to measure UHE neutrinos by detection of
Cherenkov radiation produced by $\nu_l+N\to l+N'$.
Whereas AMANDA/IceCube is optimized for detection of 
penetrating muons
resulting from $\nu_\mu+N\to\mu+N'$, RICE/ANITA are designed to detect
compact electromagnetic cascades initiated by $e^+(/e^-)$: 
$\nu_e(/{\overline\nu_e})+N\to e^\pm+N'$. As the
cascade develops, atomic electrons in the target medium are swept into the
forward-moving shower, resulting in a net charge on the shower front of
$Q_{tot}\sim E_se/4$; $E_s$ is the shower energy
in GeV.
Such cascades produce broadband Cherenkov radiation --
for $\lambda^{Cherenkov}_{E-field}>r_{Moliere}$,
the emitting region approximates a point
charge of magnitude $Q_{tot}$ and therefore emits fully
coherently; fortuitously, the field attenuation length at such
wavelengths $\ge$1 km.
One calculation finds
that, for
1 PeV$<E_{\nu_e}$,
radio detection of cascades becomes more cost-effective than
PMT-based
techniques\cite{Price96}.

The RICE experiment consists of an array of 20 in-ice dipole
receivers, deployed at depths of $100-400$ m,
and read out into digital oscilloscopes.
Calibration techniques and event reconstruction,
as well as results on the neutrino flux at earth,
are presented elsewhere\cite{Kravchenko03a}. 
An initial $\nu_e$-only
analysis based on data taken in August, 2000
has been presented elsewhere\cite{Kravchenko03b}.
Using calculations presented herein of
the expected 
radio-frequency signal strength due to an electromagnetic 
shower, the measured RICE
hardware performance,
reconstruction software and simulation\cite{Kravchenko03a},
we now report on an expanded neutrino search based on all data taken in
1999, 2000, and 2001.
We additionally use the excellent timing characteristics of the RICE
receiver array to measure the electromagnetic wave speed through the firn.

\section{Experimental Description}
\begin{figure}
\flushleft{\includegraphics[width=7.2cm]{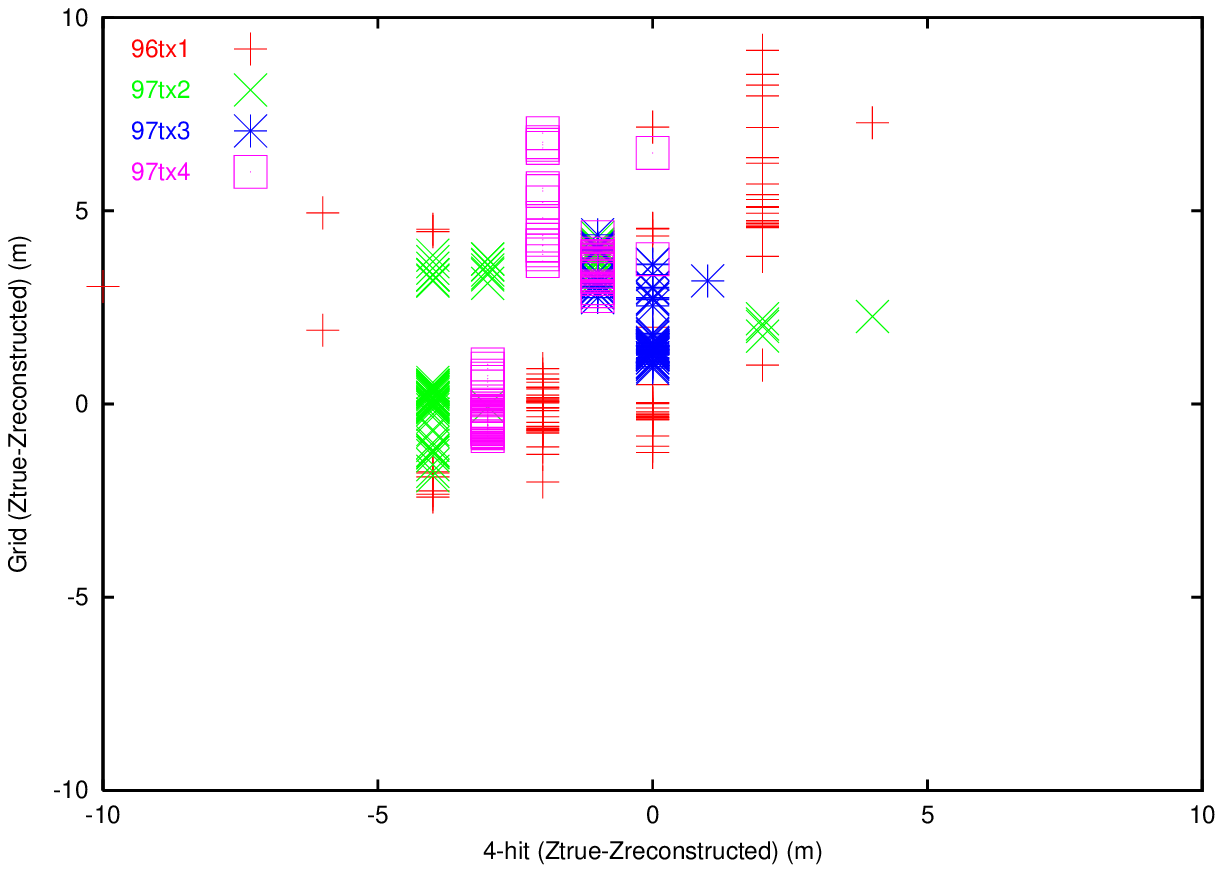}}
\vspace{-5.3cm}
\flushright{\includegraphics[width=7.3cm]{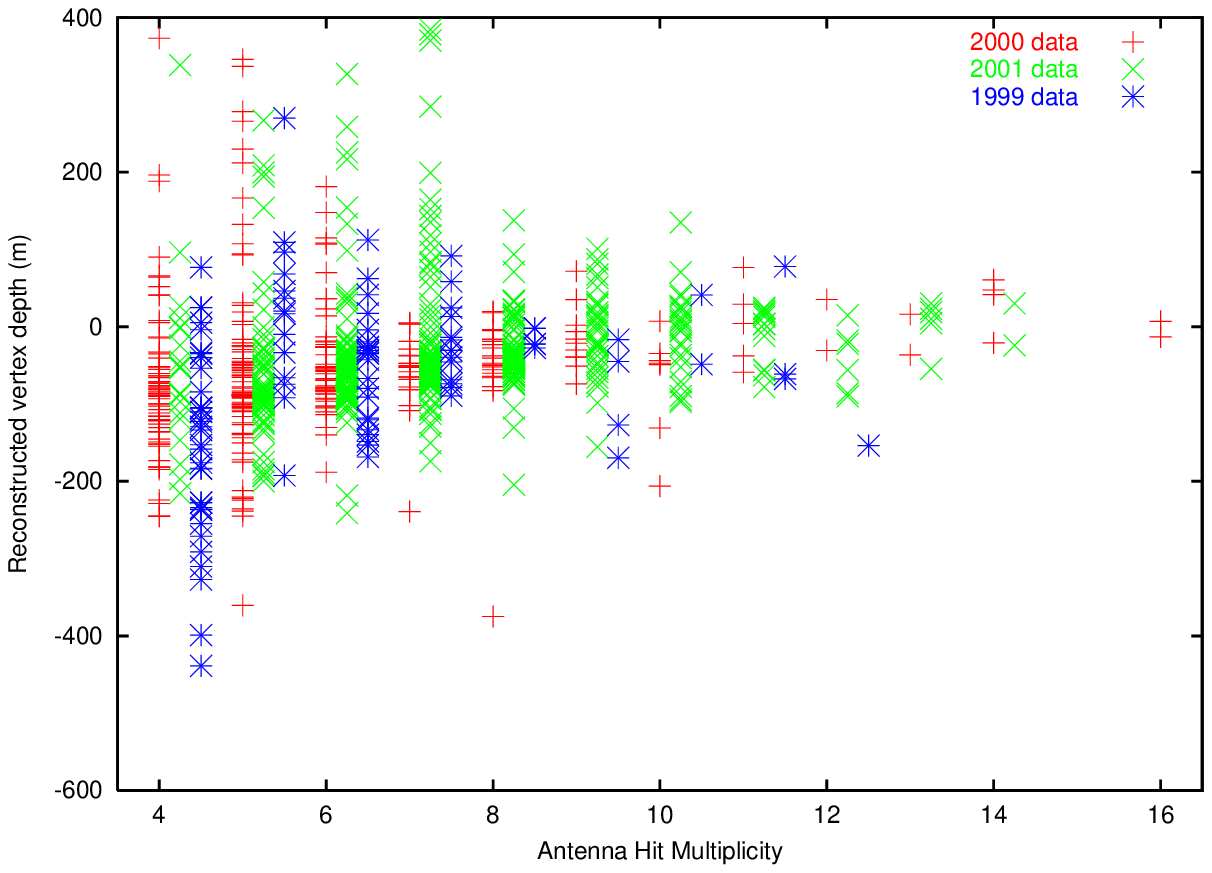}}
\vspace{.5cm}
\caption{Left (a): Deviation between true depth and reconstructed depth for
four separate transmitters, for the two source reconstruction codes;
Right (b): Raw
distribution of reconstructed z-vertex vs. hit multiplicity
for 1999, 2000, 2001 data, using
analytic vertex reconstruction algorithm. No ray-tracing corrections for
transmission through the firn have been made, which accounts for much
of the scatter in the data, nor have quality-of-fit cuts been applied.
Each point represents $\sim$50 events.}
\label{fig:alltxrecon}
\end{figure}

The RICE
experiment presently consists of a 20-channel 
array of dipole radio 
receivers (``Rx''),
scattered within a 200 m$\times$200 m$\times$200 m cube, at 100-300 m depths.
The signal from each antenna is boosted by a 36-dB in-ice amplifier, then
carried by coaxial cable to the surface observatory, where the signal is
filtered (suppressing noise below 200 MHz), re-amplified
(either 52- or 60-dB gain), and split - one copy is fed into a
CAMAC crate to form the event trigger; the other signal copy is routed into
one channel of an HP54542 digital oscilloscope.
Short-duration pulses broadcast from
under-ice transmitters provide the primary calibration signals, and
are used to verify vertex reconstruction techniques. Figure
\ref{fig:alltxrecon}a) illustrates the vertex reconstruction performance
for our calibration transmitter data (transmitters are typically 100-200 m
from receivers)
using two vertex-reconstruction algorithms. One
algorithm searches a cubic km. grid around the array for the
source point most consistent with the observed hit times; the 
second technique analytically solves for the vertex using
four-hit subcombinations of all the available hits. Typical differences
between reconstructed and known depths are of the order a few meters.
For non-calibration events, we expect reconstructed source vertices to
cluster around the surface; smearing effects due to ray tracing through 
the firn may be considerable.
Figure \ref{fig:alltxrecon}b)
displays the reconstructed source depth for our
``general'' triggers for various hit
multiplicities; source depths are observed to peak towards z=0
(consistent with surface anthropogenic activity).

\begin{figure}[htpb]
\centerline{\includegraphics[width=7.2cm]{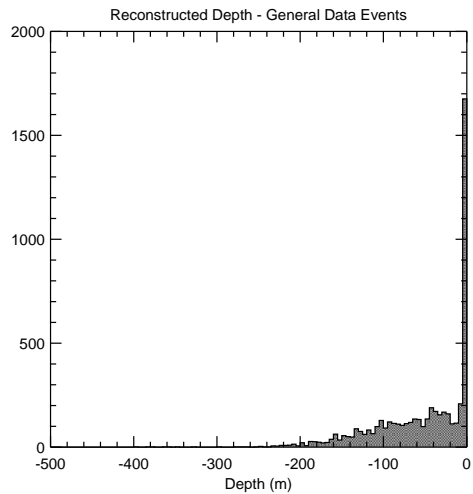}}%
\vspace{0cm}
\caption{
Reconstructed vertex depths for 5/1/03$\to$5/10/03 RICE data;
the majority of events are
consistent with a surface origin.}
\label{fig:txrxtest}
\end{figure}

With newer data,
and after applying
variable index-of-refraction corrections, the reconstructed source depth 
distribution typically sharpens, as shown in Figure \ref{fig:txrxtest}.

\section{GEANT-based calculations of RF Cherenkov signal strengths}

\subsection{Background}

Ultrahigh energy electron neutrinos from cosmological sources can be
detected from the shower created in dense media (ice e.g.) by the
secondaries in a charged current interaction ($N \nu_e \rightarrow
eX$). Shower particles travelling faster than light in the medium emit
coherent Cherenkov signals at radio frequencies, which are detected
by radio antennas buried in the medium.

We have modeled the signal contribution from the dominant ($\sim$
80\%) electromagnetic shower component using GEANT detector simulation
tools. Shower simulation and electromagnetic pulse generation from
shower particles are essentially two separate procedures in our
study. The details of our previous
studies can be found elsewhere\cite{Razzaque02}. We
report here highlights of our recent efforts.

\subsection{Shower Simulations}

We have defined a cube of ice of km scale as the target medium in our
GEANT simulations. Given an effective atomic number $Z=7.2$, an
effective mass $A=14.3$ and a density 0.92 g/cm$^3$ of the medium,
GEANT calculates all necessary parameters (radiation length,
absorption length and cross-sections, e.g.) internally.

The electromagnetic showers, in our analysis, are initiated by an $e^-$
or $\gamma$ with pre-specified momenta and position. GEANT gives
detailed particle tracking information such as interaction points, total
energy, energy lost in interaction and interaction time in output
data files. 
These data files are used later to
calculate electromagnetic pulses and to diagnose shower
properties.

We used GEANT 3.21 and, more recently, GEANT 4 to simulate
electromagnetic showers in ice. In our original 
work\cite{Razzaque02}, we used
GEANT 3.21 with default settings to generate showers, which yields
significantly less track length compared to GEANT 4 and to GEANT 3.21
with ``preferred'' settings. The reason is that, with the
default setting, electrons are stopped prematurely before reaching the
low kinetic energy threshold needed for accurate Cherenkov radio
signal emission calculations. Our updated results from GEANT
3.21 with preferred settings and GEANT 4 are in reasonable agreement
with each other and with other 
results\cite{Zas91,Zas92,Alvarez-Muniz02,Jaime02b}.  In the lower frequency
range, the signal is significantly increased compared to the signal
reported earlier. Total track lengths and particle yields at the shower
maximum are reported in Table \ref{tab:sim1} 
for an average 100 GeV $e^-$ shower
generated by GEANT 3.21 with the preferred and default settings and
GEANT 4. Also shown are results from our copy of the Zas, Halzen and
Stanev\cite{Zas91} code for comparison.

\begin{table}[t]
\caption{Track length and particle yield results from an averaged 100
GeV electron-induced shower using different Monte Carlo shower codes.
The error bars correspond to error in the mean, $s/\surd{N}$, where
$s$ is the standard deviation and $N$ is the number of showers (100
for GEANT 4 and 20 in all other cases).}
\begin{center}
\begin{tabular}{l|ccc|cc}
\hline
Shower &  \multicolumn{3}{c|}{Total track lengths} & 
\multicolumn{2}{c}{Particle yield} \\
code & Absolute & Projected & Projected
& \multicolumn{2}{c}{at shower max} \\
& ($e+p$) [m] & ($e+p$) [m] & ($e-p$) [m]
& ($e+p$) & ($e-p$) \\
\hline
G3 (preferred) & 542.74 $\pm$ 0.08 & 455.3 $\pm$ 0.2 & 125.0
$\pm$ 2.0 & 148 $\pm$ 5 & 42 $\pm$ 3 \\
G3 (default) & 389.51 $\pm$ 0.48 & 365.7 $\pm$ 0.5 & 76.3
$\pm$ 1.5 & 111 $\pm$ 7 & 20 $\pm$ 2 \\
G4 & 572.58 $\pm$ 0.04 & 466.3 $\pm$ 0.2 & 135.0 $\pm$ 0.8 & 153 $\pm$
3 & 45 $\pm$ 1 \\ 
ZHS & 642.17 $\pm$ 0.06 & 516.6 $\pm$ 0.2 & 135.2 $\pm$ 1.5 & 164
$\pm$ 6 & 44 $\pm$ 2 \\ \hline
\end{tabular}
\end{center}
\label{tab:sim1}
\end{table}

\subsection{Calculation of Radio Signal}

We have calculated the net electric field, in the Fraunhoffer limit,
by vector superposing contributions from Monte Carlo track segments of
all the charged particle using the formula:
\begin{equation}
R{\vec E}_{\omega}({\vec x}) = \frac{1}{\sqrt{2\pi}} \left(
\frac{\mu_rq}{c^2} \right) e^{ikR} e^{i\omega [t_1-(n/c) {\hat n}
\cdot {\vec r}_1]} \,{\vec v}_{\perp} \frac{e^{i\omega \delta
t(1-{\hat n} \cdot {\vec \beta}n)}-1}{1-{\hat n} \cdot {\vec \beta}n}
\label{eq:full-field}
\end{equation}
where $(t_1,{\vec r}_1)$ is the initial position of a track segment
and $\delta t$ is the time elapsed. The refractive index of the medium
is denoted by $n$ and ${\vec v}_{\perp} = - {\hat n} \times ({\hat n}
\times {\vec v})$, ${\vec v}$ being the particle velocity and ${\hat
n}$ the observer's direction. The condition $1-{\hat n} \cdot {\vec
\beta}n=0$ defines signal emission at the Cherenkov angle $\theta_c =
\cos^{-1}(1/n\beta)$. At or close to this angle
Eq. (\ref{eq:full-field}) reduces to
\begin{equation}
R{\vec E}_{\omega}({\vec x}) = \frac{i\omega}{\sqrt{2\pi}} \left(
\frac{\mu_rq}{c^2} \right) e^{ikR} e^{i\omega [t_1-(n/c) {\hat n}
\cdot {\vec r}_1]} \,{\vec v}_{\perp} \delta t \,.
\label{eq:reduced-field}
\end{equation}

\begin{figure}[t]
\begin{center}
\includegraphics[height=23.0pc]{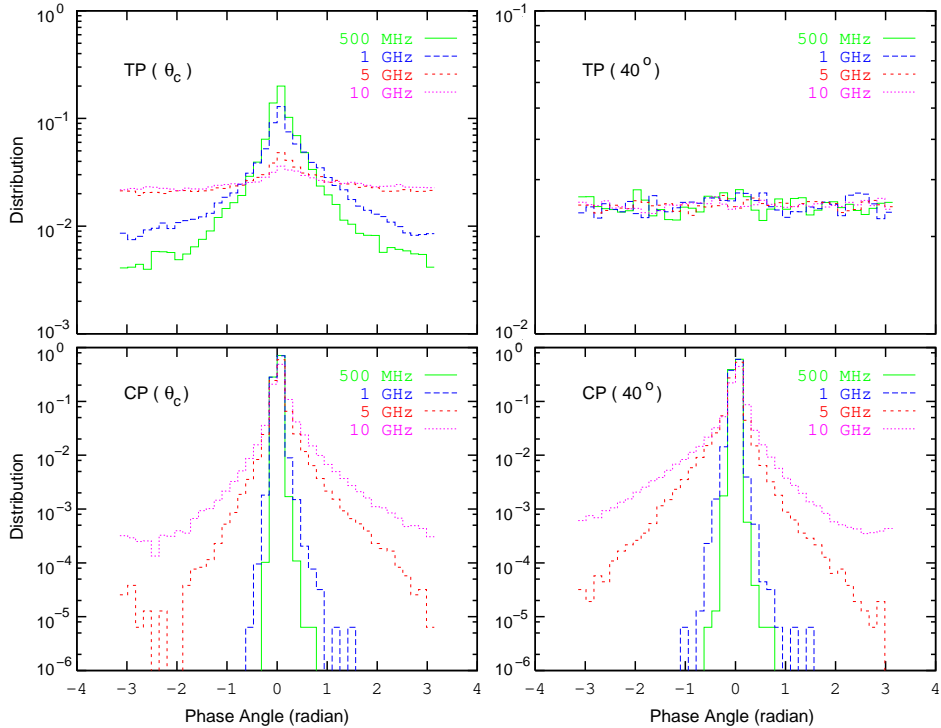}
\end{center}
\caption{Translational phase (TP) [top panels] and the Cherenkov phase
(CP) [bottom panels] distributions at the Cherenkov [left panels] and
40$^{\circ}$ angle [right panels] for a 100 GeV shower. The TP is the
determining factor in coherent signal emissions.}
\label{fig:TP-and-CP}
\end{figure}

A study of the phase angles: $\omega [t_1-(n/c) {\hat n} \cdot {\vec
r}_1]$ and $\omega \delta t(1-{\hat n} \cdot {\vec \beta}n)$ (the
translational phase (TP) and the Cherenkov phase (CP) respectively)
shows that coherent signal emission is dominated by TP (see
Fig. \ref{fig:TP-and-CP}). 
These uncorrelated phases allow one to factorize the field
equations and calculate the electric field semi-analytically by
parametrizing the shower with a form-factor. This serves as a
check of our understanding of the frequency spectrum of the electric
field at the Cherenkov angle calculated by the direct Monte Carlo
method. The form-factor itself is derived from a fit to the transverse
distribution (Fig. \ref{fig:TP-and-CP})
of the excess charge in the shower. The
agreement between the Monte Carlo and analytic spectra are good at
low frequencies ($< 10$ GHz). Fig. \ref{fig:TP-and-CP} 
shows the Monte Carlo frequency
spectrum from a 100 GeV shower.

There are several subtleties at the high frequency end of the
spectrum. The linear coherence with shower energy does not translate to
higher frequencies but some degree of coherence is retained. The
scale of coherence is influenced by the peak ($\sim 0.1$ cm) of the
transverse distribution of the excess charge in the shower which is
much smaller than the Moliere radius ($\sim 10$ cm). However,
understanding the high frequency behavior may involve addressing
questions of the role played by the high energy particles at the
beginning of the shower and the statistics of the 
signal phase relationships
from different tracks in the later stages of the shower,
for example.

\begin{figure}[t]
\begin{center}
\includegraphics[height=12.0pc]{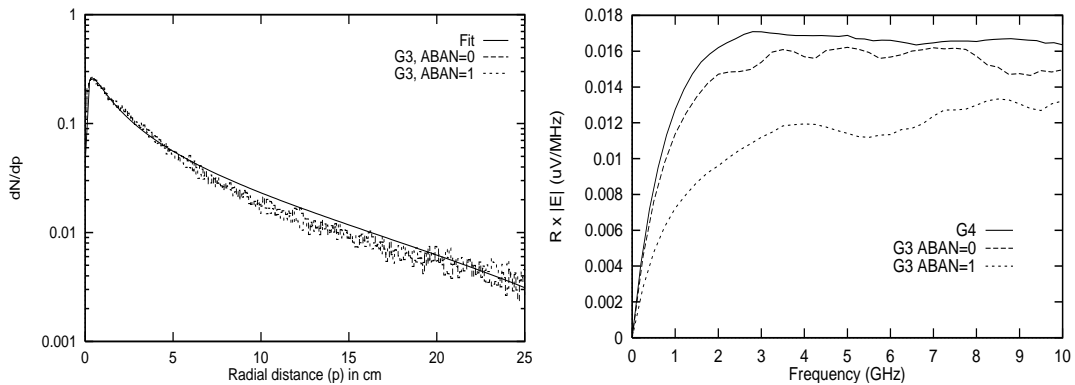}
\end{center}
\caption{Transverse excess charge distribution of a 100 GeV shower at
the maximum [left panel] and the Monte Carlo frequency spectrum [right
panel].}
\end{figure}

\subsection{Summary of Shower Simulations}

We have updated our calculations previously done with the GEANT 3.21 default
settings by using the preferred settings of GEANT 3.21 and GEANT
4. The new results yield considerably higher track lengths, number of
particles and electromagnetic signal strength. Antenna response is
directly proportional to the electric field amplitude, which in turn
affects the effective volume of the experiment.

We conclude that, for purposes of scaling radio signals from lower to
higher energy showers, one can reasonably rely on linear scaling for
frequencies below a few GHz.  Clearly there is some interesting
physics to explore in the domain above a few GHz.  This important
topic and the topic of signal contributions from the hadronic shower
component are currently under study.

\section{Limits on the neutrino flux incident on the Earth}
Although RICE has sensitivity to such physics as monopoles, topological
defects, etc., the primary aim of the RICE experiment is detection of
cosmological neutrinos.
To select neutrinos, we require
candidate events to: a) have at least 4
antenna channels registering 5$\sigma$ 
excursions in their waveforms, b) pass quality-of-vertex cuts,
c) have reconstructed vertex depths
below 150 m, c) a hit geometry
consistent(/inconsistent) 
with a conically-(/spherically-) emitting source.
Five candidate events pass all software filters;
for all five events, hand-scanning reveals at least one hit 
clearly inconsistent with the time domain antenna response expected for a true
neutrino. After elimination of such
spurious hits identified in the visual
scan, all five candidate events reconstruct near the surface.
Monte Carlo simulated waveforms (superimposed upon noise taken from data)
are used to determine event
selection efficiencies. 
Table II presents the results of our search.

\subsection{Monte Carlo Effective Volume}
The primary result of the Monte Carlo is an energy-dependent effective 
volume for detection of neutrinos.  For a given input spectrum 
and integrated livetime, the expected number of detections is then 
readily determined.  We compare this with the observed number of 
detections, and obtain an upper limit to the normalization of the input 
spectrum.
The effective volume averaged over the August 2000 exposure, for example,
is shown as the bold curve in Figure \ref{fig:eff_volume}a). 
For $E_\nu$= 100 PeV, 
$V_{eff}$ $\sim$ 1 km$^3$. 

\subsection{Results}
\begin{figure}[htpb]
\flushleft{\includegraphics[width=7.5cm]{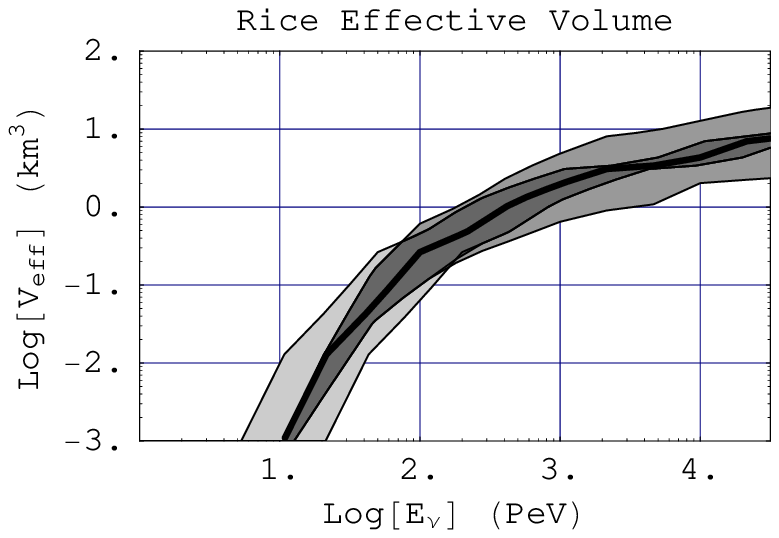}}\vspace{-6cm}
\flushright{\includegraphics[width=9.cm]{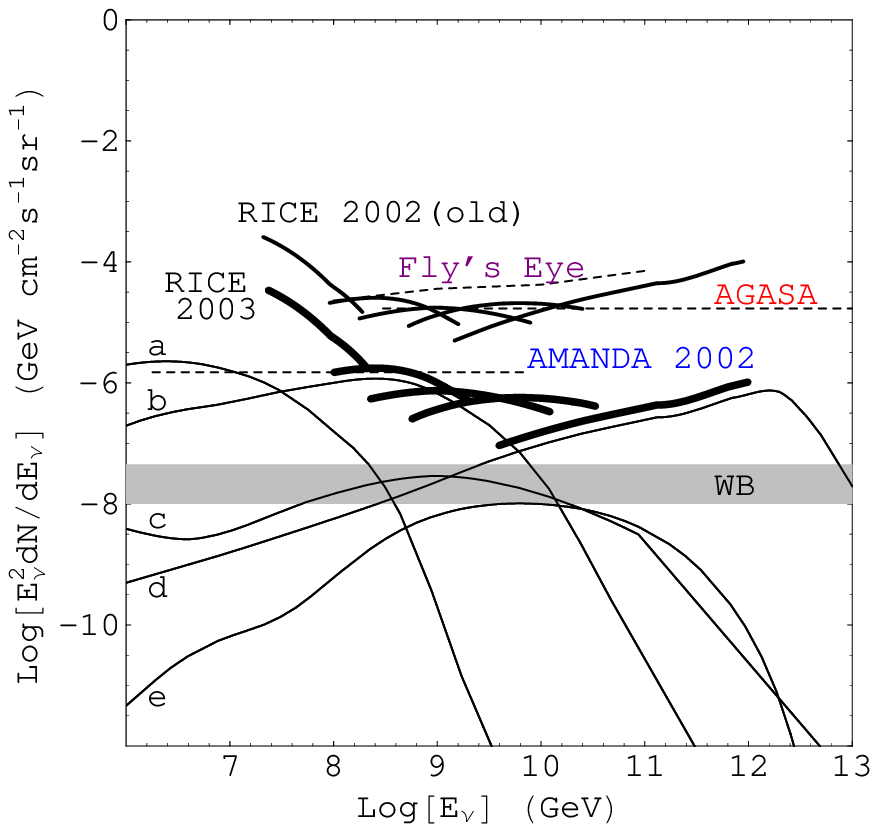}}
\caption{LEFT: Effective volume, as a function of shower energy, for the 
August, 2000 data. The nominal result corresponds to the bold curve. The 
region in light gray spans the expected response due to variation in the 
attenuation length by factors of (0.5-2.0). The region in dark gray shows
 the range due to changes in signal strength by (0.5-2.0). RIGHT: Neutrino 
flux model predictions (a)--e)) and corresponding current RICE calculated
upper limits  (95\% confidence level; thick solid), as a function of $E_\nu$.}
\label{fig:eff_volume}
\end{figure}

Given the 
known experimental
circuit gains and losses determined from an
absolute calibration\cite{Kravchenko03a}, the effective volume
$V_{eff}$ is calculated as a function of incident $E_\nu$,
as an exposure average of the detector 
configurations. The most 
important variable is the global discriminator threshold, which is adjusted to 
maintain an 
acceptable trigger rate under conditions of varying environmental noise.
Knowing the total livetime for the full
dataset (3500 hours), and based 
on observation of zero candidates,
we calculate (Figure \ref{fig:eff_volume}b) an upper limit on the
incident $\nu$ flux, as a function of incident energy.

Shown is the older upper limit from RICE (Aug., 2000 data only), as
well as the AMANDA\cite{Ahrens03},
AGASA\cite{Yoshida01} and Fly's Eye\cite{Balusaitas95} experiments (dashed). 
The predictions shown in the Figure
are:
(a)=Stecker \& Salamon\cite{Stecker96}
(b)=Protheroe\cite{Protheroe97}, 
(c)=Mannheim (A)\cite{Mannheim95}, 
(d)=Protheroe \& Stanev\cite{PS97}, 
(e)=Engel $et~al.$ GZK-model\cite{ESS01}. 
Also shown is the Waxman-Bahcall upper-limit\cite{WB99} (grey).
For a given spectral 
shape, the integrated event rate depends only on the overall 
normalization, so limits shown replicate the shape of the test spectrum 
at an amplitude corresponding to the limit. 
Each RICE upper limit segment corresponds to the neutrino energy range 
responsible for the middle 80\% of the event rate.

Improvements in the RICE upper limit over the previous limit
result from a nearly
order-of-magnitude increase in the exposure, as well as inclusion of
$\nu$-induced hadronic showers.

\begin{table}[htpb]
\begin{center}
\begin{tabular}{c|c|c} \hline
Cut imposed & Surviving Data Events  & MC events left \\ 
& (1999/2000/2001) & \\ \hline
Total triggers & 297512/111586842/3174390 & 400 \\
Passing surface veto &  12674/406867/97357 & 400\\
Passing $4\times 5\sigma_V$ cuts & 323/9001/9089 & 400 \\ 
(Z$<-$150 m)*(vertex quality) cut & 5/177/68 & 396 \\ 
Conical geometry & 0/3/2 & 378 \\ 
Passing Scanning & 0/0/0 & 376 \\ \hline
\end{tabular}
\caption{Summary of 1999-2001 data analysis.}
\end{center}
\label{tab:cuts}
\end{table}

\subsection{Prospectus}
In addition to searches for neutrinos, the RICE detector offers sensitivity
to other analyses 
(monopole detection,
studies of neutrinos coincident with GRB's and air showers, searches for
micro-black holes, etc.); results of such searches will be reported in
the future.
The 2002 and 2003 datasets comprise our highest-quality data thus far
and should offer substantial improvement over the results presented herein.
Beginning in 2004, we hope to take advantage of the scientific
opportunity presented by IceCube hole drilling to 
substantially expand the current RICE array.

\section{{\it In situ} Measurement of the Index of Refraction}
\subsection{Introduction}
Whereas neutrinos are expected to interact below
the array,
RICE RF backgrounds due to 
air showers, or above-surface 
anthropogenic sources, require recontruction of sources
viewed upwards through the firn.
Conversely, a receiver array 
deployed above the ice surface would also require ray tracing from
in-ice neutrino interactions to the above-surface detector.
This necessitates 
ray tracing the trajectories of radio waves through
a region of variable ice density and dielectric
constant, resulting in 
shorter signal propagation times and reduced
signal amplitudes (at non-zero incident angles)
compared to the case where sources are entirely
below the firn.

\subsection{Methods}
\begin{figure}
\flushleft{\includegraphics[width=6cm]{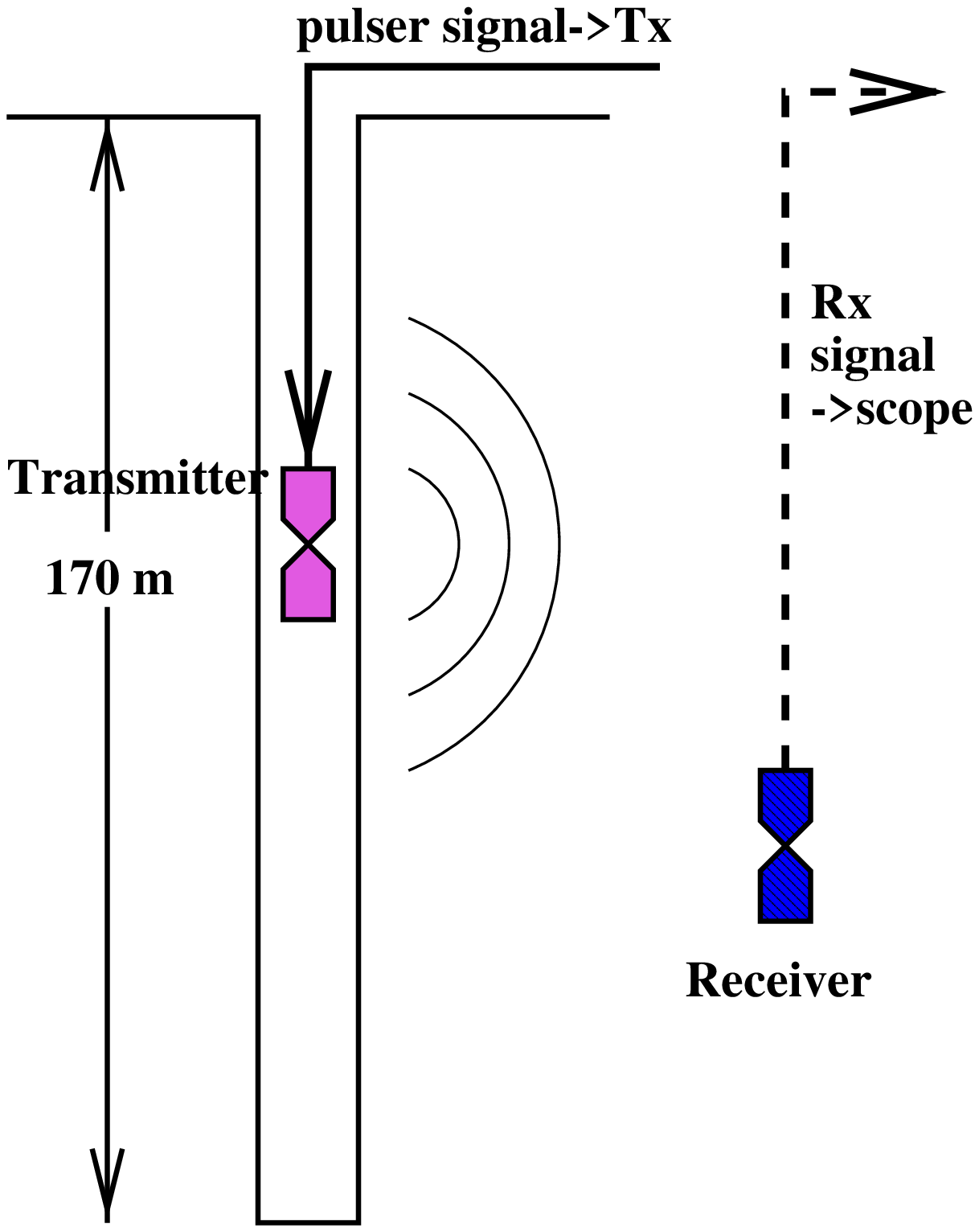}}
\vspace{-7cm}
\flushright{\includegraphics[width=8cm]{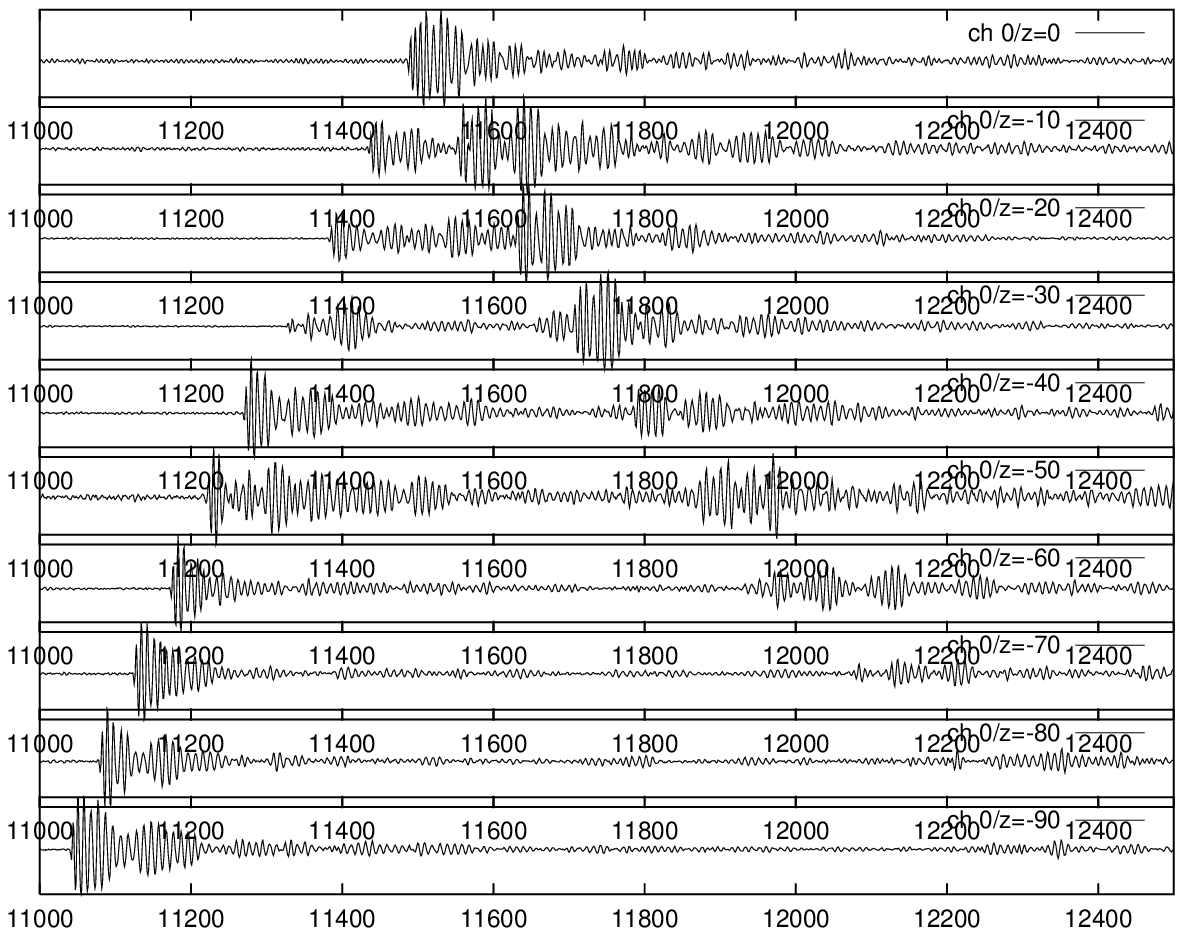}}
\vspace{1.5cm}
\caption{Left (a): Geometry of index-of-refraction
measurements. A transmitter
(Tx) is connected, via coaxial cable, to a pulse generator
in the Martin A. Pomerantz Observatory. As it is lowered into
a borehole,
the transmitter broadcasts to one of the RICE
dipole receivers (located in-ice). Right: Successive
signals recorded (at 10m depth increments) in one
receiver channel, as the transmitter is lowered into the ice. 
Note the significant after-pulsing due to signal trapped within the ice.
Horizontal units
are 0.5 ns.}
\label{fig:nVz-Tx-Rx.xfig.eps}
\end{figure}

Each RICE receiver consists of a half-wave dipole antenna, offering 
good reception over the range 0.2--1 GHz, plus a 36 dB low-noise amplifier.
The peak response of the antenna is measured to be $\sim$500 MHz in air
($\sim$300 MHz in ice), with a bandwidth ${\Delta f\over f}\sim$0.2.
An identical dipole antenna, without the amplifier, is used as the transmitter
for this measurement.
As the transmitter was slowly lowered into 
a RICE borehole,
a pulser signal was broadcast from
the transmitter (at 5-10 meter depth increments)
 to the RICE receiver array, and signal arrival times in
the receivers recorded (Figure \ref{fig:nVz-Tx-Rx.xfig.eps}). 
This afforded two measurements of the index
of refraction: 1) the index of refraction as a function of depth ($n(z)$)
could be
inferred by determining the transit time difference to a particular
receiver between successive transmitter locations 
($n=c(t_{i+1}-t_i)/(|{\vec r}_{Tx,i+1}-{\vec r}_{Rx}| - |{\vec r}_{Tx,i}-
{\vec r}_{Rx}|)$, and 2) the 
``mean'' index of refraction ($<n(\Delta z)>$),
averaged over the distance from the transmitter to any receiver could be
inferred by subtracting cable delays from the measured full
circuit (pulse generator$\to$transmitter$\to$receiver$\to$DAQ) time.
At each transmitter location, the $t_0$ of the
transmitter signal, as well as an 8.192 microsecond waveform 
(sampled at 2 GSa/s) was
recorded in an HP5452 digital oscilloscope, for each receiver.
The hit time is determined from the first $\sim 5\sigma$ excursion in each
waveform; the variation in
receiver hit time with transmitter depth is shown in Figure 
\ref{fig:nVz-Tx-Rx.xfig.eps}b.
As the transmitter approaches the (deeper) receiver, the
hit time migrates to smaller values; also evident in 
the Figure are the ``afterpulses'' corresponding to signals which
reflect off of the surface firn-air boundary, and back down to the 
buried receiver. As the transmitter is lowered to greater depths, 
the incident angle (with respect to normal surface incidence from the
under-ice transmitter location) decreases and reflection effects 
correspondingly decrease, as well.

\begin{figure}
\flushleft{\includegraphics[width=8cm]{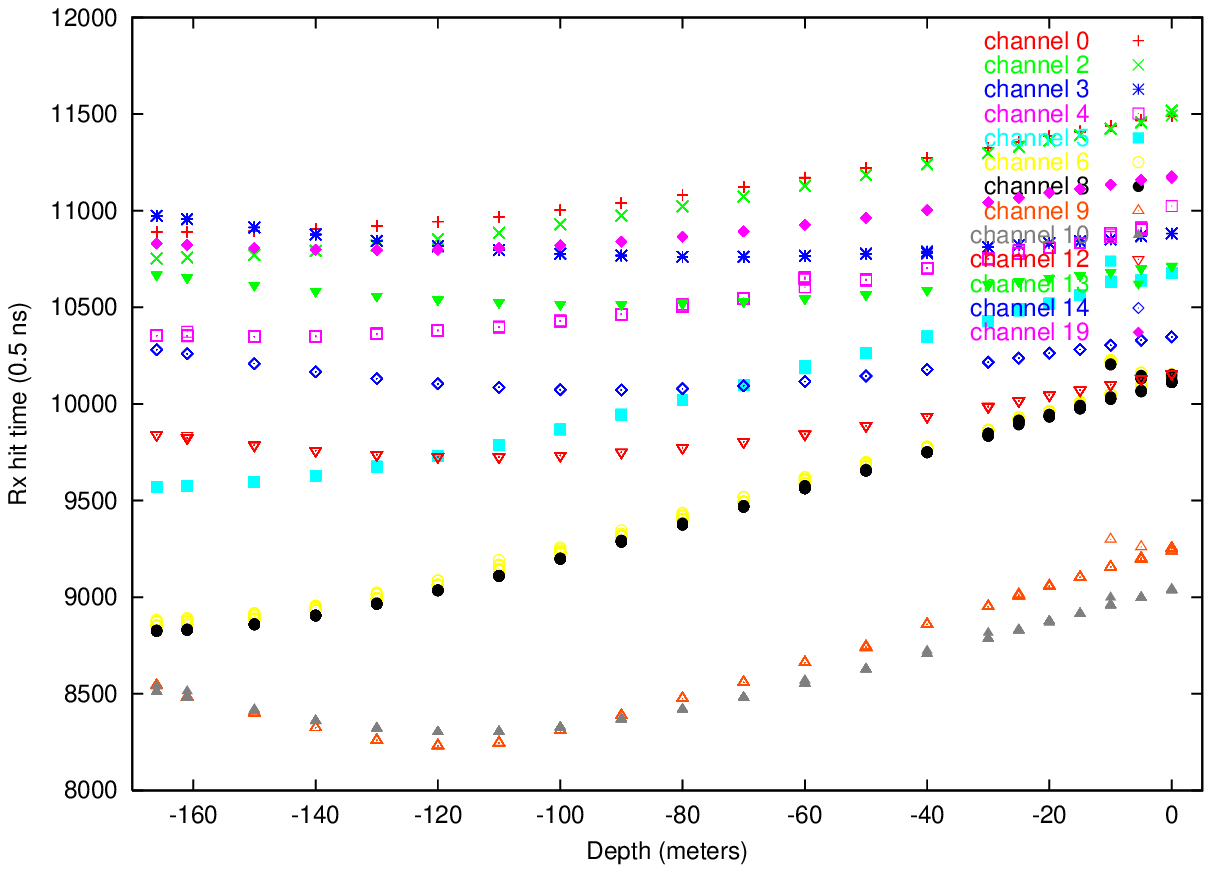}}
\vspace{-7cm}
\flushright{\includegraphics[width=7.5cm]{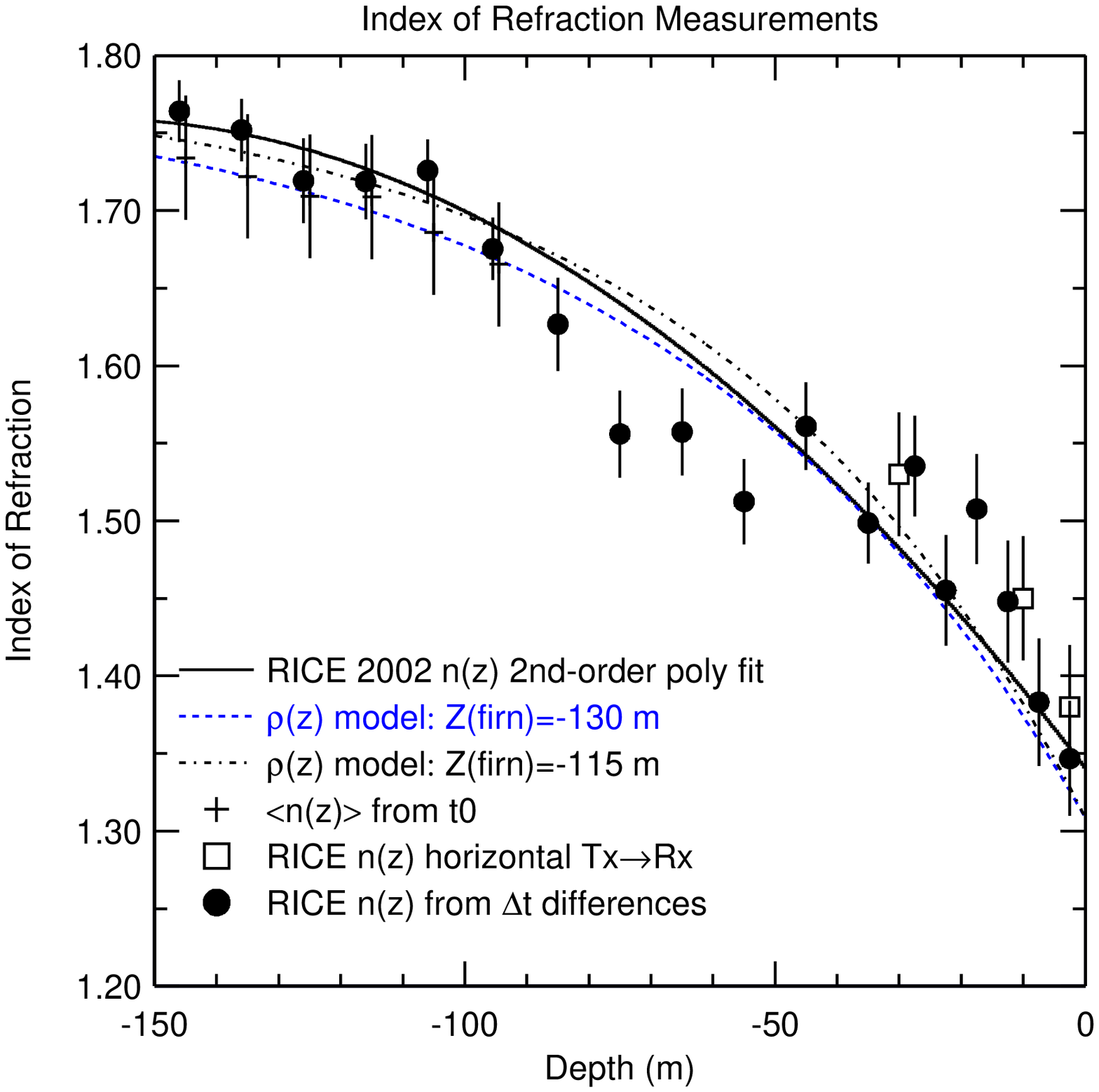}}
\caption{Left (a): Recorded hit times, for several channels, as a function of
transmitter depth. Right (b):
Final fit 
to index-of-refraction data, as described in the text.}
\label{fig:RFdraft-finalfit.eps}
\end{figure}

The local electromagnetic wave propagation velocity is
directly obtained from successive hit times such as in Figure 
\ref{fig:nVz-Tx-Rx.xfig.eps}b, 
and can therefore be
translated into an index-of-refraction profile, $n(z)$.
Figure \ref{fig:RFdraft-finalfit.eps}a) shows the
locus of hit times for several receiver channels; the local
value of n(z) can be obtained from slopes to these data points.
We have attempted to obtain an
``aggregate'' estimate of n(z) by: a) adding the raising-transmitter
plus lowering-transmitter datasets, b) adding all ``good'' data from
all possible channels, where the contribution to the final average
from each channel was weighted by geometry (favoring nearly-vertical
channels),
c) averaging over possible transmitter location uncertainties by
rebinning data and re-obtaining averages using 20 m distance
differences (rather than 10 m) between successive Tx broadcasts.
In all cases, we assumed a transmitter location uncertainty of 
0.5 m for all measurements, as well as a hit-time uncertainty of
1 ns. 
Fig. \ref{fig:RFdraft-finalfit.eps}b) shows the result of this 
procedure, and also includes data obtained by broadcasting horizontally
between a transmitter-receiver pair
being lowered simultaneously into two neighboring boreholes.
Also included are measurements derived from
the ``average'' n(z) values obtained using absolute $t_0$ measurements,
as well as comparisons with 
the predictions of the Schytt model\cite{Schytt58}, which relates index of
refraction directly to firn density $\rho(z)$, assuming that the 
ice-firn transition occurs at either z=-115 m or z=-130 m.

The RICE data points are fit to a 2nd-order polynomial, with the constraint
that the value of index of refraction at large depths approach a
constant asymptotic value. We obtained an estimate of that asymptotic
value by broadcasting from the deepest
buried RICE transmitter down to the deepest buried RICE
receiver over a distance of
233.4 m; both of these antennas are presumably well below
the firn-ice transition. The measured propagation time is 1369$\pm$8 ns,
corresponding to $n=1.764\pm0.021$.
This value is in fair agreement with the accepted value of 1.78,
as obtained by several 
measurements.
The overall error (statistical + systematic) of each data point is
estimated at 4\%.

\subsection{Check of derived n(z) using an amplitude analysis}
The variable index-of-refraction of the firn has another important
experimental consequence, due to the fact that any change in the 
index of refraction will introduce non-zero reflections
at the corresponding interface.
In any region
where there is a variation in the index of refraction,
reflection and transmission amplitudes, for both the
perpendicular and parallel components of the incident electric
field, are given by the standard 
``Fresnel coefficients'' for dielectric media\cite{Fresnel-equations}:
$r_{\bot}=-sin(\theta_i-\theta_t)/sin(\theta_i+\theta_t)$,
$r_{||}=+tan(\theta_i-\theta_t)/tan(\theta_i+\theta_t)$,
$t_{\bot}=2sin(\theta_t)cos(\theta_i)/sin(\theta_i+\theta_t)$,
$t_{||}=2sin(\theta_t)cos(\theta_i)/(sin(\theta_i+\theta_t)
cos(\theta_i-\theta_t)$,
as qualitatively illustrated in Figure
\ref{fig:nVz-Tx-Rx.xfig.eps}b.
Thus, the measured signal
strength of above-ice sources, as viewed by
below-firn receivers (or vice versa), depends on the 
degree of variation in the index of refraction.
We have used this principle to check the derived index of refraction
based on amplitudes alone.

\begin{figure}
\centerline{\includegraphics[width=12cm]{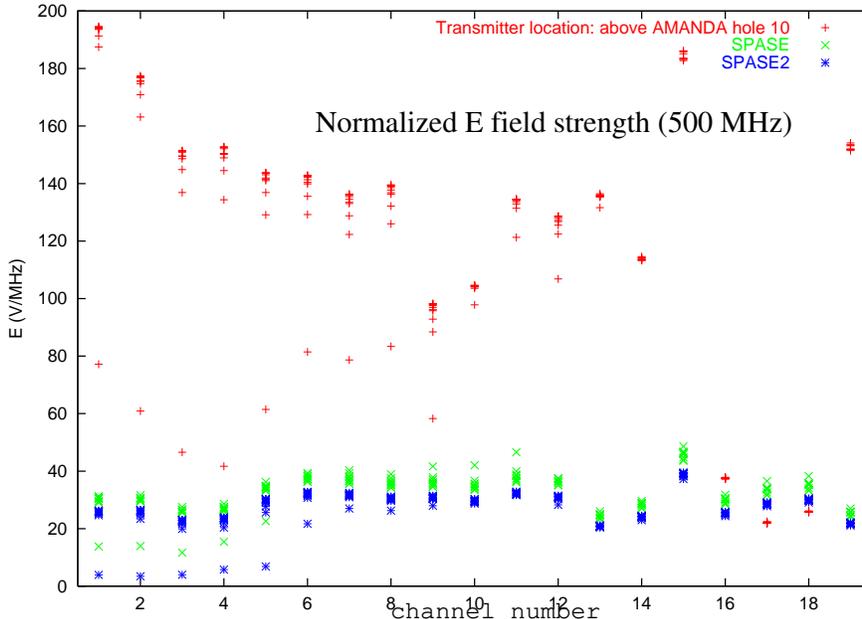}}
\caption{Received electric field strength 
at 500 MHz, normalized by channel-to-channel gain, and 
corrected for distance-to-transmitter, for data taken with: 
a) (red)
transmitter 3 meters above ice surface, and located above AMANDA hole 10;
amplitude of output signal strength = 0.3 V,
b) (green) transmitter 3 meters above ice surface, and located close to
SPASE building at South Pole, approximately 350 meters 
(measured along the surface)
from the center of the RICE receiver array;
amplitude of output signal strength = 0.3 V,
c) (blue) transmitter 3 meters above ice surface, and located close to
SPASE building at South Pole, 
amplitude of output signal strength = 1 V; received power has been
scaled by 1/3 to compare with a) and b). 
Note the agreement between the red
vs. green/blue points for channels 16/17/18, which are receivers located in
air and therefore immune to reflection effects.}
\label{fig:AM10-SPASE-XYZ-fft.eps}
\end{figure}

To investigate firn transmission effects, we have broadcast to the RICE
array from a surface point in the center of the RICE array
(on top of AMANDA hole \#10), compared to
a point having a high inclination angle relative to the RICE array 
(in the vicinity of the SPASE building, approximately 350 meters in
$r$ away from the center of the RICE array). 
Figure 
\ref{fig:AM10-SPASE-XYZ-fft.eps} shows the results of this exercise,
comparing measured electric field strengths, at the known broadcast
CW frequency, and corrected for both receiver-to-receiver channel gain,
as well as distance from transmitter to receiver.
At each transmitter location, ten waveforms were captured, and the
signal strength at 500 MHz obtained from a Fourier Transform of the
time-domain signal.
For the in-air horn antenna 
receivers, which are not subject to Fresnel losses,
there is good agreement between the distant vs. the near source 
transmitter location data. However, broadcasting at large inclination
angles from a more distant source point results in received signal
strengths smaller by an order of magnitude than what can be 
accounted for by just 1/r and gain corrections.
Our results
are qualitatively consistent with the expected Fresnel 
coefficients\cite{Fresnel-equations}. However, a rigorous comparison with 
the Fresnel equations would require consideration of several effects,
including: a) surface feature effects, b) evanescent waves which may 
propagate along the surface, and c) the true curvature of radio waves
through the firn; we have not systematically evaluated such possible
effects.
Due to the latter uncertainty,
no dipole field pattern geometry corrections have been applied --
at relatively steep angles, such dipole field pattern corrections
can be substantial. It should be noted that,
in general, 
incorporating such corrections would have
the effect of increasing the disparity between the red vs. green/blue
data points. 


\subsection{Estimate of attenuation length from Greenland data}
\begin{figure}
\flushleft{\includegraphics[width=8cm]{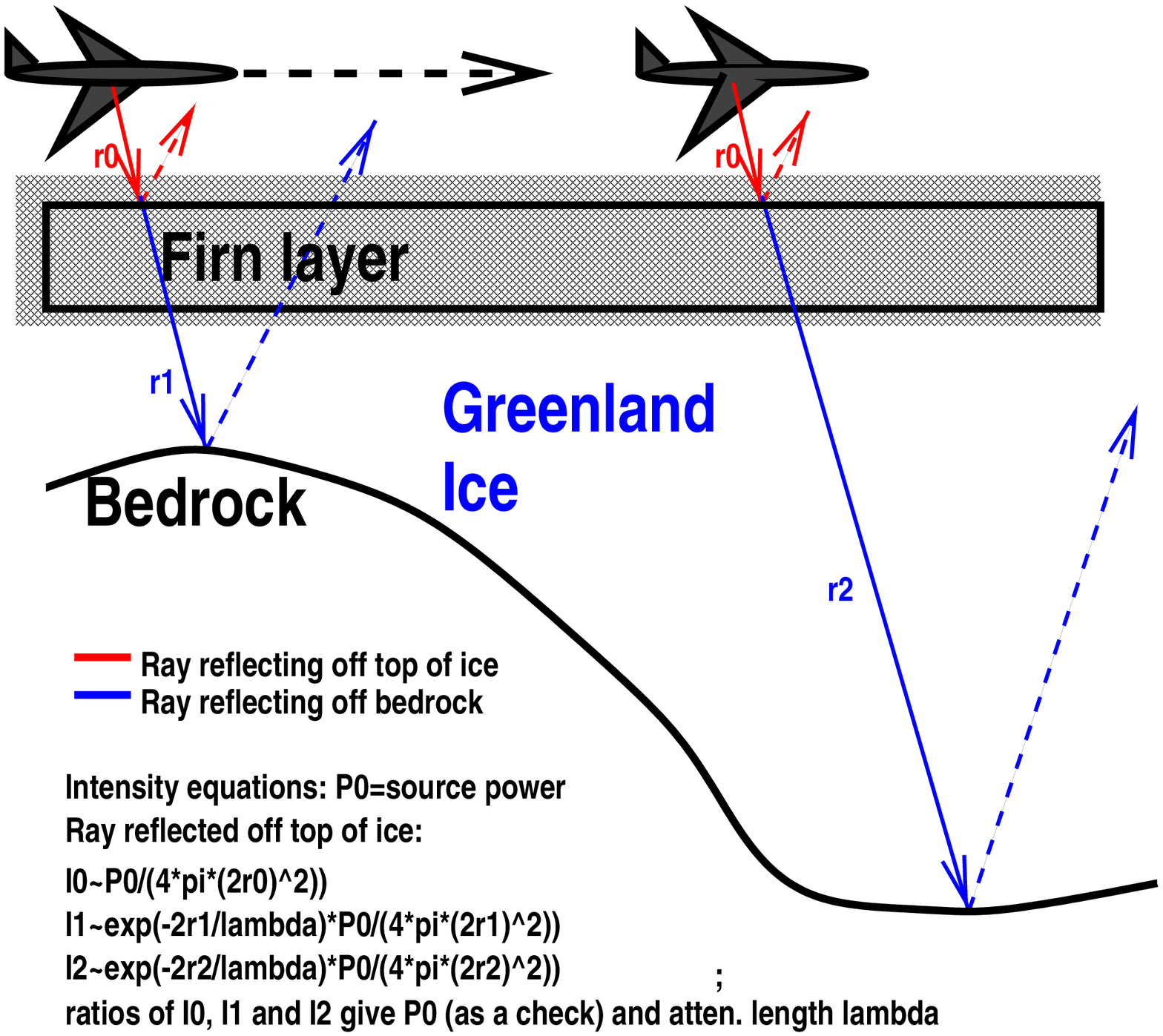}}
\vspace{-6cm}
\flushright{\includegraphics[width=8cm]{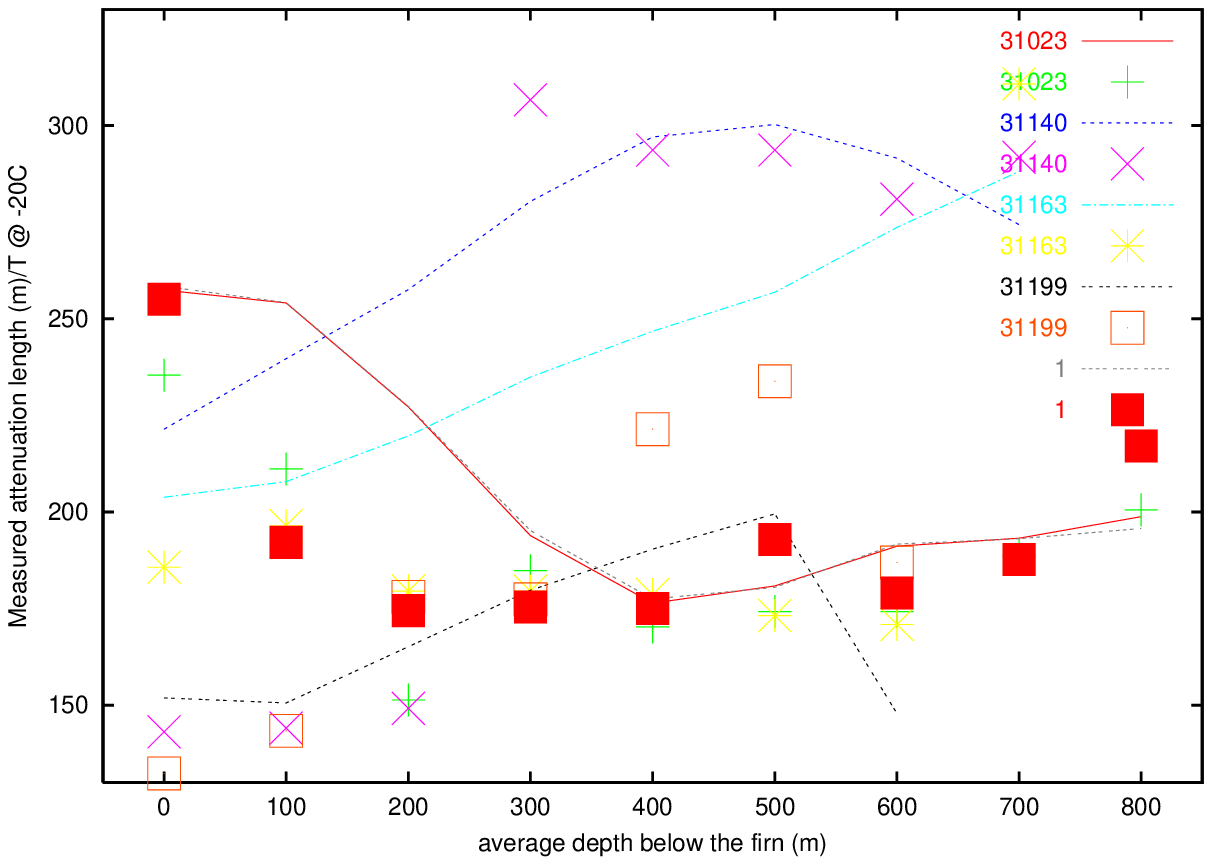}}
\vspace{0.5cm}
\caption{a) LEFT: Schematic of ground-penetrating
radar (GPR) measurements, used to
estimate the radio-frequency (f=150 MHz) attenuation
length, based on data taken by the
KU group of Gogineni {\it et al}. We take the ratio of signal amplitudes
returned by reflection off of bedrock beneath shallow ice vs. deep ice to
infer the attenuation length at intermediate depths (plotted horizontally,
and shown on the RIGHT (b)),
for several different data runs. Although the scatter from run-to-run is
large, results ($\lambda_{atten}\sim$250 m) are in general agreement with
previous estimates at this frequency and for this temperature.}
\label{fig:may31.eps}\end{figure}

We have estimated the
attenuation length in warmer (--20 C) Greenland ice, using GPR
data taken by the KU group of Gogineni {\it et al} in Greenland\cite{Gogineni}.
Here, we use the
relative ratios of ``return'' signal strengths from a transmitter, on
a low-flying plane, 
broadcasting
signals off the under-ice bedrock
(i.e., comparing the ratio of
signal strength of the bedrock echo at different
locations) to extract $\lambda_{atten}$. As indicated in Fig.
\ref{fig:may31.eps}, 
our initial estimates of the
attenuation length from these data are encouraging and in
agreement with previous measurements at these temperatures
and wavelengths. With the larger 
array size afforded by IceCube, a direct {\it in situ} 
measurement using RICE receivers and transmitters should be
straightforward.

\section{Acknowledgments}
We gratefully acknowledge the support of The Research
Corporation, the NSF Office of Polar
Programs under grant \#OPP-0085119, as well as the University of
Kansas Undergraduate Research Awards.
The RICE experiment would not have been possible without the generous
logistical
and material support of the AMANDA Collaboration.
We gratefully acknowledge additional
support 
from
the KU General Research Fund and the KU Research Development Fund,
the New Zealand Marsden Foundation, 
and the  
Research Corporation. 

\end{document}